\documentclass[a4paper, 11pt]{article}

\usepackage[english]{babel}

\usepackage[a4paper,top=2cm,bottom=2cm,left=3cm,right=3cm,marginparwidth=1.75cm]{geometry}

\usepackage{amsmath, amssymb, amsfonts, amstext}
\usepackage{graphicx}
\usepackage{natbib}
\usepackage[colorlinks=true, allcolors=blue]{hyperref}
\usepackage{aas_macros}
\usepackage{bm}
\usepackage{xcolor}
\usepackage{subcaption}
\usepackage{graphicx}

\newcommand{\comment}[1]{} 

\title{\textbf{\LARGE Big Bang Nucleosynthesis with \bm{$f(R)$} gravity scalarons and astrophysical consequences}}
\author{\small
	Abhijit Talukdar$^{1}$\footnote{\href{mailto:abhijittalukdar@gauhati.ac.in}{abhijittalukdar@gauhati.ac.in}} \ 
	and Sanjeev Kalita,$^{1}$\footnote{\href{mailto:sanjeev@gauhati.ac.in}{sanjeev@gauhati.ac.in}}
	\\
	\small $^{1}$Department of Physics, Gauhati University, Jalukbari, 781014, India\\
}

\date{}

\begin{document}
\maketitle

\begin{abstract}
	$f(R)$ gravity is one of the serious alternatives of general relativity having {a} large range of astronomical consequences. In this work, we study Big Bang Nucleosynthesis (BBN) in $f(R)$ gravity theory. We consider modification to gravity due to the existence of primordial black holes in the radiation era which introduce additional degrees of freedom known as scalarons. We calculate the light element abundances by using the BBN code \texttt{PArthENoPE}. It is found that for a range of scalaron mass $(2.2-3.5) \times 10^4$ eV, the abundance of lithium is lowered by $3-4$ times the value predicted by general relativistic BBN which is a level desired to address the cosmological lithium problem. For the above scalaron mass range {the} helium abundance is within {the} observed bound. However, {the} deuterium abundance is found to be increased by $3-6$ times the observed primordial abundance. It calls for {a} high efficiency of stellar formation and evolution processes for destruction of primordial deuterium which is suggested as possible in scalaron gravity. A novel relation between scalaron mass and black hole mass has been used to show that the above scalaron mass range corresponds to primordial black holes of sub-planetary mass ($\sim 10^{19}$ g) serving as one of the {potential candidates of non-baryonic dark matter}. We infer Big Bang equivalence of power law $f(R)$ gravity with primordial black holes that are detectable with upcoming gravitational wave detectors.
\end{abstract}



\section{Introduction}\label{sec1}

With only six free parameters -- (1) cosmic baryon density  $(\rho_b)$, (2) dark matter density ($\rho_\text{DM}$), (3) Hubble parameter ($H_0$), (4) spectral index ($n_s$) measuring departure from scale invariance of primordial density perturbations, (5) amplitude of primordial density perturbation $(\delta_i)$ and (6) optical depth ($\tau$) for scattering of the CMB photons by intergalactic medium during cosmic reionization, the standard $\Lambda$CDM cosmology has successfully explained cosmic expansion history and origin of {the} large scale structures of the universe \citep{2003RvMP...75..559P, 1990Natur.348..705E, 2015PNAS..11212246P}. It is based on flat spatial sections of the Friedmann-Lemaitre-Robertson-Walker geometry. It predicts that the universe started with a Hot Big Bang (HBB) singularity. One of the remarkable achievements of the HBB model is {its ability to predict abundances} of the light elements, deuterium (D), helium ($^3$He, $^4$He) and lithium ($^7$Li) which formed through {a sequence of capture reactions between primordial protons and neutrons  \citep{PhysRevLett.16.410,1966ApJ...146..542P,1967ApJ...148....3W}}. After 1 s of the Big Bang when the universe cooled down to $\sim10^{10}$ K, the rate of interactions keeping protons and neutrons in equilibrium fell below the cosmic expansion rate \citep{Weinberg2008}. This is the ‘freeze-out’ epoch. As the {universe cooled down to} $\sim10^9$ K nuclear reactions started locking free protons and neutrons which were frozen in a fixed proportion. The Big Bang Nucleosynthesis (BBN) of the $\Lambda$CDM cosmology assumes standard laboratory physics of nuclear reactions, three neutrino species and no additional relativistic degrees of freedom \citep{2015PNAS..11212246P}.

Cosmic inflation \citep{PhysRevD.23.347, Linde:1983gd} driven by a primordial scalar field has greatly supported the $\Lambda$CDM cosmology. Quantum fluctuation of energy of the scalar field generates seed density fluctuations. These fluctuations were stretched and later amplified by gravity leading to cosmic structures. Primordial black holes (PBHs) are the earliest structures formed by large amplitude random density {fluctuations} \citep{1968ApJ...151..431M, 1974MNRAS.168..399C}. {Formation of the PBHs depends on details of the inflation model. It might not be a natural process in the standard cold inflationary scenario. However, production of PBHs is found to be quite natural in the warm inflation scenario where the scalar field has non-negligible interactions with the pre-existing matter fields \citep{2020JCAP...09..042A, 2021JCAP...12..052B}. \cite{2022PhLB..83537510C} advocated for production of high mass PBHs by considering shift-symmetry-protected natural inflation under warm inflationary paradigm.} A PBH of mass $M$ forms if cosmological density of the early universe, $\rho=3/32\pi Gt^2$ achieves density required to form a black hole, $3M/{4\pi R_s^3}$ where $R_s =2GM/c^2$ is the Schwarzschild radius. This implies linear growth of the PBH mass, $M \sim c^3t/G$ . \cite{1971MNRAS.152...75H} advocated for PBHs with masses $10 ^{-5}$ g upward. This minimum bound arises from classical nature of gravity -- that the Schwarzschild radius of a black hole must not be smaller than the Planck length ($l_\text{Pl}\sim 10^{-33}$ cm) -- the scale {of the universe when it was} $10 ^{-43} $ s old.  These PBHs, however, undergo Hawking evaporation. The evaporation time-scale is shorter for smaller black holes \citep{HAWKING1974}. Extremely low mass PBHs evaporate long before the onset of BBN. PBHs with masses around $10^{13}\text{ g}$ evaporated by the epoch of matter-radiation decoupling. PBHs with mass above $10^{15}\text{ g}$ survive the Hubble time. These PBHs formed around $10^{-24}$ s after the Big Bang. This is long before the era of BBN which constrains the amount of baryonic matter in the universe. Therefore, these PBHs act as potential candidates of non-baryonic dark matter. Mass windows of PBHs which can contribute to the mass budget of non-baryonic dark matter have been succinctly reported in \cite{2021JPhG...48d3001G}. It has been found that {PBHs with mass} $10^{17}-10^{22}$ g are eligible {candidates} for non-baryonic dark matter.

The standard cosmology is, however, not free from problems. Cosmological singularity at the beginning of the universe demands quantum corrections to {General Relativity} (GR). This {calls for} alternative theories of gravity. The $\Lambda$CDM cosmology relies on unknown physics of the cosmological constant and {Cold Dark Matter} (CDM). There are alternatives to the $\Lambda$CDM paradigm. These include time varying dark energy components \citep{1988PhRvD..37.3406R,PhysRevLett.80.1582,PhysRevLett.82.896} within GR and modification of GR on the cosmological scales \citep{Capozziello:2002rd, 2003PhRvD..68l3512N,Starobinsky2007}. Distinguishing between these two is {one of} the major {goals} of ongoing Dark Energy Survey \citep{2005astro.ph.10346T} and the upcoming Euclid mission \citep{2011arXiv1110.3193L}.  In addition the {standard} cosmology has recently experienced a sequence of serious tensions. The notable ones are the Hubble tension \citep{2021ApJ...919...16F} and the $\sigma_8$ tension \citep{2019ApJ...876..143B, 2013A&A...555A..30B}. These call for serious alternatives to the paradigm \citep{2021CQGra..38o3001D}. 

The BBN is plagued by cosmological lithium problem. Whereas abundances of D, $^4$He and $^3$He match with those observed in intergalactic medium, interstellar medium and the solar system \citep{2003ApJS..149....1K, 1995ApJ...451..335L, 1996Sci...272..846N, 1995ApJS...97...49O, 1994ApJ...435..647I, 2002ApJ...565..668P}, the $^7$Li abundance has an anomaly. The predicted relative abundance of lithium ($^7$Li/H) is $(4.68\pm0.67)\times{10}^{-10}$ \citep{RevModPhys.88.015004} while the observed relative abundance is reported as $^7$Li/H=$(1.58^{+0.35}_{-0.28}) \times 10^{-10}$ \citep{2010A&A...522A..26S}. This discrepancy of a factor of $3-4$ is known as the lithium problem. Neither new observation nor nuclear physics consideration of the BBN process has been able to remove this discrepancy \citep{2021PhLB..81236008A}. This casts doubt in one or more of the following -- the success of the BBN, thermal history of the universe and cosmology of the early universe. Available attempts to address the problem include variation of fundamental physics during BBN \citep{2002PhRvD..65l3511I, 2006ApJ...637...38L, 2010PhLB..683..114B, 2017ApJ...834..165H, 2019PhRvD..99l1302V}.

Once the physics of the BBN such as cosmic expansion rate and nuclear reaction cross sections are fixed, the predicted abundances of the light elements depend only on one parameter -- the baryon-to-photon ratio $(\eta=n_B/n_\gamma)$. Assuming nuclear and other non-gravitational physics to be accurate during the BBN era, we investigate effect of modified gravity theory on the cosmological abundances of light elements. Here we consider {the} additional scalar degree of freedom of $f(R)$ modified gravity theories as an extra energy density component in the Friedmann evolution of the radiation era. Along with the baryon-to-photon ratio this serves as input parameter in predicting modified elemental abundances. 

The study of modified theories of gravity in the BBN era was initiated by \cite{1968ApJ...152....1D}. He proposed that astronomical test of the Brans-Dicke scalar-tensor theory of gravity is possible through determination of helium abundance in Population II stars. This was found to be possible as the scalar field energy density of the modified theory enhances the expansion rate so much that helium production becomes forbidden during the BBN era. Abundances of all other light elements get affected as this theory changes the rate of cooling of the hot universe. Result is shift of the freeze-out epoch and hence a change in the frozen proportion of neutrons and protons. \cite{1968Ap&SS...2..155G} and \cite{1978MNRAS.184..677B} studied constraint on temporal variation of Newton’s constant, $G$ embedded in the Brans-Dicke theory through observed abundances of D and $^4$He. In recent times viable cosmological models in alternative gravitation theories such as Hořava-Lifshitz theory, $f(Q,T)$ gravity theory have been proposed through consistency with BBN \citep{2022IJMPA..3750017B, 2010JCAP...01..013D}. Observed abundances of D and $^4$He are used by \cite{2015PhRvD..91j4023K} to constrain $f(R)$ gravity theory. These theories are used for building models of the accelerated cosmic expansion.    

Black holes provide with astrophysical environments for testing alternative theories of gravity \citep{2012PhRvD..85l4004B, 2017PhRvL.118u1101H}. $f(R)$ theories constitute a class of alternatives to GR where the left hand side of Einstein’s field equations are altered \citep{2011PhR...505...59N}. They are often used for explaining the observed cosmic acceleration \citep{2003astro.ph..3041C} and the dark matter phenomena \citep{2007MNRAS.375.1423C} without incorporating exotic fields or particles \citep{Nojiri:2008nt}. It has been shown \citep{2018ApJ...855...70K, 2020ApJ...893...31K} that gravitational correction (polynomial functions of Ricci scalar, $R$) to quantum fluctuation in empty space around black holes naturally produces such alteration of the theory of gravity. The empty space field equations are derived from the modified Einstein-Hilbert action,
\begin{equation}
	A_{E-H}=\frac{c^4}{16\pi G}\int{f(R)\sqrt{\left|\text{det}g_{\alpha\beta}\right|}d^4x}
\end{equation}

Here $\text{det}g_{\alpha\beta}$ represents determinant of the spacetime metric tensor and $f(R)$ is a function of the Ricci scalar curvature, $R$. The scalar mode of gravity present in these theories is described by the scalaron field, $\psi=df(R)/dR$. Astronomical consequences of $f(R)$ gravity scalarons have been extensively studied by using stellar orbits near the Galactic Center black hole and the shadow size of the black hole \citep{2020ApJ...893...31K, 2021ApJ...909..189K, 2023IJMPD..3250021P, 2024ApJ...964..127P}.
Mass of the scalaron is computed from ultraviolet (UV) and infrared (IR) cut off scales of curvature corrected vacuum fluctuations near black holes \citep{2020ApJ...893...31K} as,
\begin{equation}
	m_\psi=\frac{2\pi}{\lambda_\text{IR} \lambda_\text{UV}}\sqrt{\frac{\lambda_\text{IR}^2-\lambda_\text{UV}^2}{12\ln (\lambda_\text{IR}/\lambda_\text{UV})}} 
\end{equation}
where, $\lambda_\text{UV}$ and $\lambda_\text{IR}$ represents UV and IR cutoff scales respectively. The UV scale is chosen as the Schwarzschild radius, $\lambda_\text{UV}=R_\text{s}=2GM/c^2$. The IR scale is chosen as the one which presents vacuum thermal energy density corresponding to Gibbons-Hawking temperature of black holes and is written as  $\lambda_\text{IR}=(hc/a_B)^{1/4}T_\text{H}^{-1}$, where, $T_\text{H}$ is the Hawking temperature of a black hole (see \cite{2020ApJ...893...31K} for details of such calculations). The scalaron mass ($m_\psi$) is found to be inversely proportional to the black hole mass $(M)$ and in natural units $c=1{=h}$ it is given by \citep{2024JCAP...02..019T},
\begin{equation}\label{eq1.2}
	m_\psi\approx{10}^{-10} \text{ eV} \left( \frac{\text{M}_\odot}{M}\right) 
\end{equation}

In a recent study it has been reported that this naïve relation ensures existence of stationary and asymptotically flat black hole solutions in $f(R)$ gravity \citep{2024ApJ...964..127P}.

In this work we consider scalarons associated with PBHs in the BBN era for which the density parameter of scalarons falls below closure density $(\rho_\text{closure} \sim \rho_r \sim T^4)$. Existence of PBHs from sub-planetary mass ($10^{17}-10^{22}$ g) to 100$\text{ M}_\odot$ in the early universe is found to be possible in hybrid inflationary cosmology with isocurvature density perturbation \citep{2022NuPhB.98415968A}. First we prove that scalarons in the early universe {constitute an} additional dark energy component with constant density and negative pressure in a power law $f(R)$ gravity theory. The elemental abundances have been estimated by adding scalaron dark density in the BBN code \texttt{PArthENoPE} \citep{2022CoPhC.27108205G} which uses the latest data on nuclear reaction rates \citep{2021JCAP...04..020P}. It has been found that scalarons with masses around 10 keV are eligible to reduce the lithium abundance by a factor $\sim3$ relative to the GR based Big Bang prediction without harming abundances of helium inferred from several independent astronomical probes. However, the deuterium abundance for these scalarons has been found to be elevated relative to the observed bounds. 

The paper is organised as follows. Section~\ref{sec2} demonstrates that $f(R)$ gravity scalarons act as a constant density dark energy component. In Section~\ref{sec3}, the cosmological density parameter of scalarons is expressed in terms of scalaron mass. Section~\ref{sec4} contains effect of scalarons on {the} BBN abundances of light elements. Section~\ref{sec5} contains results and discussions. Section~\ref{sec6} concludes.

\section{Scalaron as a dark energy component}\label{sec2}

We consider $f(R)$ gravity field equation \citep{amendola_tsujikawa_2010},
\begin{equation}
	f'(R)R_{\alpha\beta}-\frac{f(R)}{2}g_{\alpha\beta}=\nabla_\alpha \nabla_\beta f'(R)-(\nabla^\mu \nabla_\mu f'(R))g_{\alpha\beta}.
\end{equation}

{We write the above equation in Einstein form},
\begin{equation}
	R_{\alpha\beta}-\frac{R}{2}g_{\alpha\beta}=T_{\alpha\beta},
\end{equation}
where, $T_{\alpha\beta}$ is the energy momentum tensor of the scalaron field and is given by,
\begin{equation}\label{2c}
	T_{\alpha\beta}=\frac{\nabla_\alpha \nabla_\beta \psi}{\psi}-\frac{\Box\psi}{\psi}g_{\alpha\beta}+\frac{g_{\alpha\beta}}{2}\left(\frac{f(R)}{\psi}-R\right).
\end{equation}

Energy momentum tensor for a fluid is given by,
\begin{equation}\label{2d}
	T_{\alpha\beta}=(\rho+p)u_\alpha u_\beta-pg_{\alpha\beta},
\end{equation}
where, $\rho$, $p$ and $u$ represent density, pressure and velocity of the fluid respectively.

Comparing equation~(\ref{2c}) and (\ref{2d}), we obtain pressure and density of the scalaron field as,
\begin{subequations}
	\begin{align}
		p_\psi&=\frac{2\Box\psi+R\psi-f(R)}{2\psi} \label{2e_1},\\
		&\rho_\psi=\frac{f(R)-R\psi}{2\psi} \label{2e_2}. 
	\end{align}
\end{subequations}

This gives the equation of state of scalarons as,
\begin{equation}\label{2f}
	\omega_\psi=\frac{p_\psi}{\rho_\psi}=\frac{{2\Box\psi}}{{f(R)-R\psi}}-1.
\end{equation}

We call this as ``scalaron fluid''.

\subsection{Massless scalarons as cosmological constant}

It is seen that the scalaron field acts like a dark fluid similar to the cosmological constant, $\omega=-1$, if the scalaron field $\psi$ satisfies the following two conditions simultaneously,
\begin{subequations}\label{2g}
	\begin{align}
		\Box\psi&=0 \label{g1}, \\
		f(R) &\neq R\psi \label{g2}.
	\end{align}
\end{subequations}

Equation~(\ref{g1}) is the Klein-Gordon equation of a massless scalaron field.

In flat FLRW cosmology, the metric tensor components are: $g_{\mu\nu}=\text{dia}(1,\text{ } -a^2{(t)},\text{ } -a^2{(t)}r^2,\text{ } -a^2(t)r^2sin^{2}\theta)$. With this metric the first condition becomes,
\begin{equation}\label{}
	\Box\psi=\Ddot{\psi}+3H\Dot{\psi}=0,
\end{equation}
where, the dot represents derivative with respect to cosmic time and $H=\dot{a}/a$ is the Hubble parameter. This gives the following solution,
\begin{equation}
	\Dot{\psi} \sim a^{-3}.
\end{equation}

Writing $\Dot{\psi}=(d\psi/da)(da/dt)$ and assuming a power law expansion of the universe as $a(t) \sim t^\alpha\text{ }(\alpha > 0)$, we obtain,
\begin{equation}
	\frac{d\psi}{da} \sim a^{({1}/{\alpha})-4}.
\end{equation}

This gives the scalaron field as a function of scale factor,
\begin{equation}\label{13}
	\psi(a) =\frac{df(R)}{dR} \sim a^{({1}/{\alpha})-3}.
\end{equation}

Ricci scalar is considered as reciprocal of the square of the length scale $ct$. Therefore,
\begin{equation}
	R \sim a^{-2/\alpha}.
\end{equation}

This gives, ${df(R)}/{dR} \sim R^{{(3\alpha-1)}/{2}}$. The outcome is a power law $f(R)$ gravity theory,
\begin{equation}
	f(R) \sim R^{{(3\alpha+1)}/{2}} \sim R^m,
\end{equation}
where, $m=(3\alpha+1)/2>0$.

The second condition (equation~(\ref{g2})) can be parametrized as,
\begin{equation}
	f(R)=\chi R \psi, \text{ $\chi \neq 1$}.
\end{equation}
Therefore, 
\begin{equation}
	\frac{df(R)}{f(R)}=\frac{1}{\chi}\frac{dR}{R}.
\end{equation}
This has a power law gravity solution,
\begin{equation}
	f(R) \sim R^{1/\chi}.
\end{equation}

The massless scalaron field in power law $f(R)$ gravity behaves as the cosmological constant. We generalize this result for massive scalarons in the following manner.
\subsection{Massive scalarons as general dark energy fluid}

With the help of equation~(\ref{2e_2}), equation~(\ref{2f}) can be written as,
\begin{equation}\label{2_2_a}
	\omega_\psi=\frac{\frac{\Box\psi}{\psi}}{\rho_\psi}-1.
\end{equation}

For a constant equation of state different from the cosmological constant ($\omega\neq -1$) and for constant dark density, we must satisfy the following two conditions simultaneously,
\begin{subequations}
	\begin{align}
		\frac{\Box \psi}{\psi}=c_1' \label{eq20a}, \\
		\rho_\psi=c_1 \label{eq20b},
	\end{align}
\end{subequations}
where, $c_1'$ and $c_1$ are constants. Equations~(\ref{eq20a}) and~(\ref{eq20b}) are to be satisfied separately so that the scalaron field acts as a massive Klein-Gordon field and presents a constant dark density component. In flat FLRW cosmology, the Klein-Gordon equation (equation~(\ref{eq20a})) for massive scalaron field $\psi$ is given as,
\begin{equation}\label{}
	\Ddot{\psi}+3H\Dot{\psi}-c_1'\psi=0.
\end{equation}

For the power law expansion model of the universe $a(t) \sim t^\alpha\text{ }(\alpha > 0)$, we follow the procedure similar to the massless case and obtain the differential equation for evolution of the scalaron field,
\begin{equation}
	t^{2\alpha-2}\left( \frac{d^2\psi}{da^2}\right) +4t^{\alpha-2}\left( \frac{d\psi}{da}\right)=c_1'\frac{df(R)}{dR}. 
\end{equation}

For $R\sim t^{-2}$, the above equation can be written as,
\begin{equation}\label{eq23}
	R^m\left( \frac{d^2\psi}{da^2}\right) +4R^n\left( \frac{d\psi}{da}\right)=c_1'\frac{df(R)}{dR}, 
\end{equation}
where, $m=1-\alpha$ and $n=(2-\alpha)/2$. We consider the following two cases.

\textit{Case I: }Here we consider a slow variation of the scalaron field and make a Taylor expansion of $\psi(a)$ around an initial epoch $a=a_i$. It gives,
\begin{equation}\label{25}
	\psi(a)\approx \psi(a_i)+\left( \frac{d\psi}{da}\right)_i(a-a_i)+ \frac{1}{2}\left( \frac{d^2\psi}{da^2}\right)_i(a-a_i)^2.
\end{equation}

{Using equation~(\ref{25}) in (\ref{eq23}) we get the following condition},
\begin{equation}\label{eq25}
	R^m c_3+4 R^n (c_2+c_3a)=c_1'\frac{df(R)}{dR},
\end{equation}
where,
\begin{subequations}
	\begin{align}
		c_2=\left( \frac{d\psi}{da}\right) _i, \\
		c_3=\left( \frac{d^2\psi}{da^2}\right) _i.
	\end{align}
\end{subequations}

Cosmic scale factor is related to the Ricci scalar as $a\sim R^{-\alpha/2}$. This gives equation~(\ref{eq25}) as,
\begin{equation}
	\frac{df(R)}{dR}\sim R^m + R^n +R^q,
\end{equation}
where, $q=n-(\alpha/2)$.

This produces power law gravity,
\begin{equation}
	f(R) \sim R^s.
\end{equation}

\textit{Case II:} Here we consider power law evolution of the scalaron field, $\psi(a)\sim a^u$, similar to the massless case (see equation~(\ref{13})). The power law variation of the scalaron field is motivated by inflationary and quintessence scalar field models \citep{1988PhRvD..37.3406R}. Hence, equation~(\ref{eq23}) takes the form,
\begin{equation}
	\frac{df(R)}{dR} \sim R^m a^{u-2}+ R^n a^{u-1}.
\end{equation}

This condition also permits a power law gravity $f(R)\sim R^w$.

Therefore, it is a general consequence that both massless and massive scalarons behave as a constant dark density fluid $(\omega=-1,\text{ } \omega=\text{constant}\neq-1, \text{ }\rho_\psi=
\text{constant})$ in a power law gravity theory.

Hence, scalarons in a power law $f(R)$ gravity theory are eligible to act as a constant dark density fluid. It is to be noted that power law $f(R)$ gravity has important impact in cosmology. It has been used as a viable model for cosmic inflation. The most notable $f(R)$ gravity model for inflation is the quadratic gravity theory, $f(R)=R+\alpha R^2$ \citep{1980PhLB...91...99S}. It has been proved to be a reliable model of inflation after the release of the Planck 2018 data \citep{2020A&A...641A..10P}. Power law $f(R)$ gravity is found to connect early time (inflation) and late time (dark energy) accelerated expansion \citep{2023PhLB..84337988O}. \cite{2015PhRvD..91j4023K} used BBN abundances to constrain the power law $f(R)$ gravity.

\section{Scalaron density in a flat universe}\label{sec3}

Considering scalarons as additional constant dark density component in the flat FLRW universe, the Friedmann equation for expansion rate of the radiation dominated universe is written as,
\begin{equation}\label{eq3.1}
	H^2=\frac{8 \pi G}{3}({\rho_r+\rho_{m_\psi}}),
\end{equation}
where, $\rho_r$ and $\rho_{m_\psi}$ are the mass densities of radiation and scalaron respectively.

We simplify Friedmann equation (equation~(\ref{eq3.1})) as,
\begin{equation}\label{eq2}
	H^2=H^2_\text{GR} m_r.
\end{equation}

Here, $H_\text{GR}=\sqrt{(8\pi G\rho_r/3)}$ is the Hubble expansion rate in GR and $m_r=1+({\rho_{m_{\psi}}}/{\rho_r})$. The critical density of the universe in scalaron gravity is written as,
\begin{equation}\label{eq3}
	\rho_c^{f(R)}=\rho_c^\text{GR} m_r,
\end{equation}
where, $\rho_c^{f(R)}={3H^2}/{8 \pi G}$ and $\rho_c^\text{GR}$ are {the} critical densities in $f(R)$ and GR scenarios respectively. The dimensionless cosmological density parameter of scalarons is written as,
\begin{equation}\label{eq4}
	\Omega_{m_\psi}=\frac{\rho_{m_\psi}}{\rho_c^{f(R)}}.
\end{equation}

In this study we take scalarons as non-relativistic {degree of freedom}. It has been found that for non-relativistic scalarons the scalaron density parameter is dependant on scalaron mass (see below) and therefore, it is possible to constrain scalaron mass by using the density parameter in the BBN code. Assuming scalarons to be in thermal equilibrium with radiation and applying Bose-Einstein statistics in the radiation era, the scalaron mass density has been obtained as \citep{2024JCAP...02..019T},
\begin{equation}
	\rho_{m_\psi}=\frac{g}{\sqrt{2}\pi^2\hbar^3}(m_\psi k_B T)^{3/2}\zeta(3/2)\Gamma(3/2),
\end{equation}
where, $m_\psi$ and $T$ represent {scalaron mass} and temperature respectively. $\zeta(\nu)$ and $\Gamma(\nu)$ represent Riemann zeta function and gamma function for a number $\nu$ respectively. It has been found {earlier} that {the} scalaron density parameter is independent of scalaron mass in the relativistic case \citep{2024JCAP...02..019T}. Therefore, the BBN calculation cannot be used for constraining scalaron mass for the relativistic case. The BBN abundances and interpretations in this study are confined to non-relativistic scalarons. The radiation density is given by,
\begin{equation}
	\rho_r=Na_\text{B} T^4/c^2,
\end{equation}
where, $N$ is the number of effective relativistic degrees of freedom and is expressed as,
\begin{equation}
	N=\frac{g_\text{boson}}{2}+\left( \frac{7}{16}\right) g_\text{fermions}\left(\frac{4}{11}\right)^{4/3} \approx1.68.
\end{equation}

Density parameter for scalarons is, therefore, obtained as,
\begin{equation}\label{eq5}
	\Omega_{m_\psi}\approx\frac{1}{1+(2\times10^{-99})\text{ }T^{5/2} m_\psi^{-5/2}}
\end{equation}

We have considered scalaron mass {in the} range $(1-10^5)\text{ eV}$. This mass range is enclosed by $(10^{-16}-10^5)$ eV which was obtained earlier from the observed shift of freeze-out temperature \citep{2024JCAP...02..019T}. In this work the higher end of scalaron mass is considered for obtaining the lower bound on the PBH mass (see equation~(\ref{eq1.2})).  The {scalaron} density parameter is estimated at the freeze-out temperature $(T \sim 9\times 10^9 \text{ K})$. For the scalaron mass range considered in this work, the scalaron density parameter is found to be in the range $10^{-16}-10^{-3}$.

\section{Scalarons and BBN abundances}\label{sec4}

Several astrophysical probes have been used to estimate the abundances of the light elements. {Helium ($^4$He) abundance is a reliable probe of new physics of the early universe. It is because the primordial abundance of $^4$He depends on the amount of neutron fraction at the freeze-out epoch. This amount is sensitive to expansion rate of the universe which in turn is governed by departure from standard physics. Therefore, accurate measurement of $^4$He abundance is necessary for understanding physics of the early universe \citep{RevModPhys.88.015004}.} Primordial $^4$He abundance is obtained from the spectroscopic study of {the} metal poor H{\sc~ii} region NGC 346 in the Small Magellanic Cloud as $Y_p=0.2451\pm 0.0026\  {(1\sigma)}$ \citep{2019ApJ...876...98V}. Another independent measurement of $^4$He was reported from a pristine intergalactic gas cloud at cosmological redshift $z_\text{abs}=1.724$ (containing about less than 30\% less metal content than the most metal poor H{\sc~ii} regions) as $Y_p=0.250^{+0.033}_{-0.025}\  {(1\sigma)}$ \citep{2018NatAs...2..957C}. 

Deuterium abundance is an accurate estimator of cosmological baryon density \citep{1973ApJ...179..343W}. {Any observed abundances of deuterium gives upper bound on the baryon-to-photon ratio, $\eta$.} Measurements of D/H ratio by using high redshift absorbing clouds along the line of sight to distant quasars were reported earlier \citep{1996Natur.381..207T, 1998ApJ...499..699B}. For more than two decades D/H ratio was measured by using high resolution spectrographs on large telescopes employed for observing absorption lines in the quasar spectra \citep{1996Natur.381..207T, 2012A&A...542L..33N, 2014ApJ...781...31C, 2018ApJ...855..102C}. The most precise measurement of deuterium abundance by using low hydrogen column density {of one of the} absorber clouds at redshift $z_{abs}=3.572$ {in the line of sight of quasar PKS1937–101} was reported by \cite{2017MNRAS.468.3239R}. It is found as D/H=$(2.62 \pm 0.05) \times 10^{-5} \  {(1\sigma)}$. Incorporating all prior measurements \cite{2024MNRAS.528.4068K} reported the best estimate of deuterium abundance as $\text{D/H}=(2.533\pm 0.024)\times 10^{-5}  \  {(1\sigma)}$. Another observation of primordial D/H is taken from an absorption system along the line of sight of quasar {Q1243+307} located at redshift $z_{abs}=2.52564$ which is found out to be D/H=$(2.527\pm 0.030)\times 10^{-5} \  {(1\sigma)}$ \citep{2018ApJ...855..102C}. Since there is no known process of creation of deuterium nuclei, these abundances are considered as the primordial ones produced in the BBN.{ There is astrophysics associated with the observed deuterium abundance. There is a trend of monotonic decrease of deuterium abundance over time. This indicates that galactic chemical evolution has affected the interpretation of local measurements D/H \citep{RevModPhys.88.015004}.}

{Metal poor halo stars of the Galaxy are potential probes of primordial $^7$Li abundance. Halo main sequence stars which are cooler than $T_\text{eff}=6000$ K destroy much of the primordial $^7$Li as they have thick convection zone. However, the hot halo stars with $T_\text{eff}>6000$ K have thin convection zone due to which $^7$Li does not go down to layers where it gets depleted. Consequently these stars preserve primordial $^7$Li. \cite{1982A&A...115..357S} observed that $^7$Li abundance remains flat for very low metallicity stars ([Fe/H]=-3 to -1.5). The $^7$Li abundance for halo stars with this metallicity is known as the ``Spite plateau'' \citep{RevModPhys.88.015004} and provides with the primordial abundance. It has been found that $^7$Li abundance is less than the standard prediction of the Big Bang model. Observed abundance of lithium is taken as  $^7$Li/H=$(1.58^{+0.35}_{-0.28}) \times 10^{-10}\ {(2-3\sigma)}$ \citep{2010A&A...522A..26S}. This is derived from observations of metal poor halo stars.}

\begin{figure}[ht!]
	\centering
	\includegraphics[width=0.5\columnwidth]{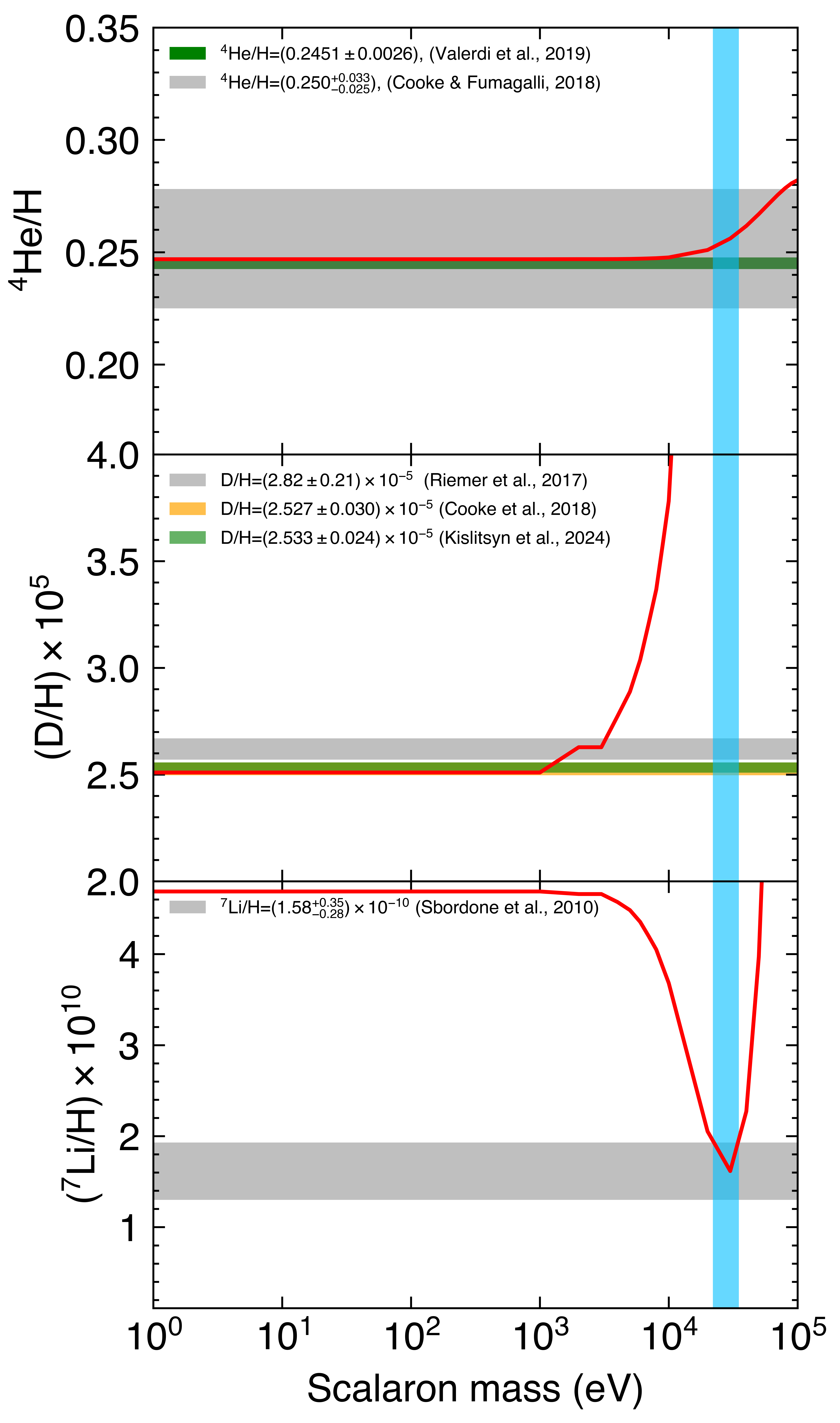}
	\includegraphics[width=0.49\columnwidth]{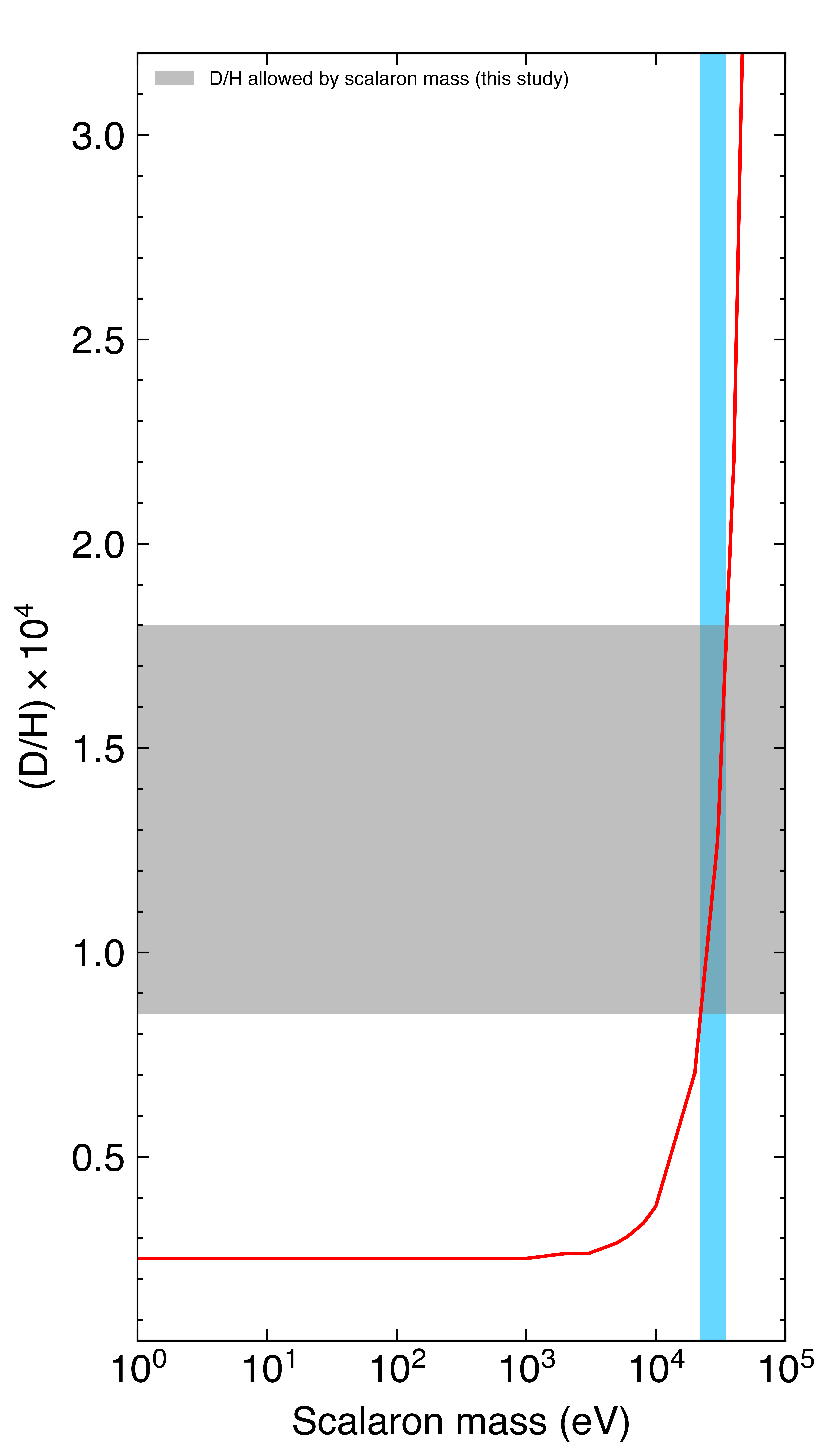}
	\caption{Left: Abundances of light elements with scalaron mass evaluated at $T\sim 9 \times 10^9$ K. Right: D/H abundance for scalaron mass showing the enhancement. \label{fig1}}
\end{figure}

Theoretical description of the BBN process is pretty well known \citep{1967ApJ...148....3W, 1969ApJS...18..247W, 1973ApJ...179..343W, 1992STIN...9225163K, 1993ApJS...85..219S}. A number of public and private codes like \texttt{PArthENoPE} \citep{Pisanti2008, Consiglio2018, 2022CoPhC.27108205G}, \texttt{AlterBBN} \citep{Arbey2012, Arbey2020}, \texttt{PRIMAT} \citep{Pitrou2018} have been in development for detailed calculation of BBN abundances. In this work, we have chosen the latest version of the code \texttt{PArthENoPE} which is \texttt{PArthENoPE 3.0}. This code allows users to calculate the abundances of elements in the standard BBN as well {as} in alternative cosmologies. \texttt{PArthENoPE} takes different parameters of BBN {such as} baryon-to-photon ratio ($\eta_{10}=\eta/10^{-10}$), number of additional neutrino species ($\Delta N_\nu$), neutron $\beta$-decay time-scale ($\tau_n$), dimensionless dark density component ($\rho_\Lambda$ {in the code which is ${\Omega_{m_\psi}}$ in our case}) as input and computes the detailed evolution of elemental abundance of light elements from nuclear statistical equilibrium condition. To execute this, we have considered $\Omega_{m_\psi}$ as additional dark density parameter in the Friedmann-Lemaitre expansion law. This component increases the expansion rate and consequently affects the BBN abundances.

To calculate the abundances using the BBN code the neutron lifetime is taken as, $\tau_n=(879.4\pm 0.6)$ s \citep{2020PTEP.2020h3C01P}. Baryon density determined by Planck CMB observation is taken as $\Omega_b h^2=(0.02242 \pm 0.00014)$ \citep{2020A&A...641A...6P}. The corresponding baryon-to-photon ratio is $\eta_{10}=(6.13815 \pm 0.04239)$ (see \cite{2006JCAP...10..016S} and \cite{ 2018ApJ...855..102C} for the conversion formula between $\Omega_b h^2$ and $\eta_{10}$). The variation of abundances D/H, $^4$He/H and $^7$Li/H as a function of scalaron mass $m_\psi$ is shown in Fig.~\ref{fig1} (Left). The results of this study are discussed below.

\section{Results and Discussions}\label{sec5}

In this work we have investigated the effect of $f(R)$ gravity scalarons on the abundances of D, $^4$He and $^7$Li using the BBN code \texttt{PArthENoPE}. In flat FLRW cosmology, power law $f(R)$ gravity theory is found to present scalarons as constant density dark energy component including the cosmological constant for massless scalarons. We take dark density of the scalarons in the Friedmann equation of the radiation era and study the variation of light element abundances with scalaron mass at the freeze-out epoch.

It has been found that scalarons are eligible to bring down lithium abundance by a factor of $3-4$ from that of standard BBN. It addresses the lithium problem. The mass range of scalarons for which this occurs is $(2.2-3.5)\times 10^4$ eV. Whereas helium abundance is found to be compatible with observed bounds, deuterium abundance for these scalarons is found to be elevated with respect to the observed abundance by a factor of $3-6$.

The primordial helium abundance falls in the range $0.250-0.257$ and is compatible with the {abundance estimated} from observation of metal poor H{\sc~ii} regions and pristine intergalactic gas clouds. The deuterium abundance shows interesting behaviour with further prospects. The calculated abundance is higher than the observed abundance by a factor of $3-6$ (D/H $\sim 8.5\times10^{-5}- 1.8 \times 10^{-4}$, see Fig.~\ref{fig1} (Right)).  In late 1990s there were reports of high primordial D/H ratio ($\sim 10^{-4}$) \citep{1997seim.proc..345W, 1996ApJ...459L...1R, 1996NuPhS..51...71F, 1997Natur.385..137S}. A low deuterium abundance ($\sim 2.6\times10^{-5}$) measured with absorbers of quasar spectra is result of observations in specific locations in space and is considered as standard primordial abundance. A discordant D/H ratio is thought to be an indicator of cosmological inhomogeneity \citep{1997seim.proc..345W}. \cite{2000PhR...333..409T} critically examined this discordance with the possibility of hydrogen contamination in the absorbers which yields high D/H ratio. Hydrogen absorption looks much like that of deuterium giving rise to apparent enhancement of D/H ratio in some absorbers. Depletion of early high deuterium is a possibility. Although global depletion by a factor of $4$ may seem unlikely \citep{2000PhR...333..409T}, stellar formation and evolution is a well appreciated source of reduction of D/H. Deuterium is extremely fragile and gets destroyed at or above $10^6$ K. Stellar atmospheres (corona) and supernova explosions easily exceed this threshold. As more and more stars form and evolve D/H ratio goes down in the clouds enriched by stellar ejecta. It is suggested that stellar processes can destroy deuterium only by a maximum of 1\% \citep{2000PhR...333..409T}. The problem may not be closed yet. It is believed that star formation history and model of galactic chemical evolution may need re-evaluation after the new era of the James Webb Space Telescope (JWST) which has observed dusty galaxies in the high redshift  universe. JADES (JWST Advanced Deep Extragalactic Survey) JWST/NIRSpec spectroscopy has reported very rarely seen N{\sc~iii}] $\lambda$ 1748 line, which suggests unusually high $N/O$ abundance in galaxy GN-z11 at redshift ${z}>10$ \citep{2023A&A...677A..88B} possibly indicating the formation of high metallicity stars. It calls for high star formation efficiency in the early universe which is not expected in the standard $\Lambda$CDM cosmology \citep{2023NatAs...7..731B, 2023NatAs...7..611R}. Rapid star formation process indicates destruction of a high primordial deuterium abundance \citep{1998ApJ...498..226T}. 

Presence of PBHs can accelerate gravitational collapse process relative to that expected in the standard $\Lambda$CDM model \citep{Carr:2020xqk}. It leads to early growth of massive galaxies. It has been found that PBHs with $10^4 \text{ M}_\odot$ can induce large star formation efficiency $(\epsilon=32\%)$ thereby accounting for existence of high redshift galaxies with high stellar mass content, $10^{10} – 10^{11} \text{  M}_\odot$ \citep{2024A&A...685L...8C, 2022MNRAS.514.2376L}. It naturally ameliorates the recently realized puzzle of JWST massive high redshift galaxies with large stellar mass \citep{2023arXiv230701457G}. Scalarons associated with the PBHs provide a natural mean to accelerate gravitational instabilities. They provide with an additional attractive gravitational force with Yukawa potential, known as the `scalaron fifth force' \citep{2018ApJ...855...70K, 2020ApJ...893...31K}. Equation~(\ref{eq1.2}) suggests that as PBHs grow in mass scalaron mass decreases. A $10^4 \text{ M}_\odot$ PBH corresponds to $10^{-14}$ eV scalarons. A $10^6\text{ M}_\odot$ black hole corresponds to $10^{-16}$ eV scalarons. These scalarons present stellar scale range $(\sim 0.0014 – 0.14$ au) of the Yukawa fifth force. Therefore, an accelerated growth of stellar mass can be naturally expected with scalarons. This may likely to provide with a seed for realistic stellar mechanism of destroying primordial deuterium. Taking the uncertainty of cosmic star formation history we infer that if primordial deuterium abundance is high, lithium abundance is low by a factor of $3-4$ relative to the standard BBN and $^4$He abundance is compatible with all existing measurements then $f(R)$ gravity theory with scalaron masses $(2.2-3.5)\times 10^4$ eV can have a Big Bang equivalence.

According to the relation between scalaron mass and black hole mass (see equation~(\ref{eq1.2})) our derived scalarons correspond to primordial black holes of masses around $10^{19}$ g. These are sub-planetary scales including the asteroids (smaller than Ceres) and some moons of the solar system. They formed around $10^{-20}$ s after the Big Bang and survive the Hubble time against Hawking evaporation. They are eligible for being candidates of non-baryonic dark matter. This mass falls within the range $10^{17}-10^{22}$ g which is unconstrained and hence is open as a window that provides all of the dark matter density in the universe \citep{2013PhRvD..87b3507C, 2013PhRvD..87l3524C}. PBHs within the above range are produced through isocurvature density perturbations in specific inflationary models and thereby contribute to Scalar Induced Gravitational Wave (SIGW) signals with frequencies ranging from nHz to kHz \citep{2022NuPhB.98415968A}. \cite{2024PhRvD.109d3537G} calculated the SIGW spectrum of these PBHs and found that GW detectors such as eLISA, BBO and DECIGO are capable of detecting these signals. Detection of PBHs with $10^{19}$ g, therefore, will hint towards a Big Bang equivalence of $f(R)$ gravity theory. Growth of PBHs with mass $10^{19}$ g (this work) to massive PBHs naturally calls for a dynamical scalaron field with cosmological coupling of scalaron mass. This calls for further investigation of scalarons in the early universe. We infer that $f(R)$ gravity with a Big Bang equivalence allows sub-planetary mass PBH dark matter.     

\section{Conclusion}\label{sec6}
A power law $f(R)$ gravity theory allows its scalar gravitational mode to act as a constant density dark energy fluid. This dark (scalaron) density is found to be compatible with the BBN abundance of helium. It's presence reduces the lithium abundance to the level desired to address the cosmological lithium problem. However, deuterium abundance is found to increase beyond observed bounds for the scalaron mass range addressing the lithium problem. If a high deuterium abundance in the early universe is treated as an open problem we infer that a power law gravity theory may call for exceptionally high efficiency of stellar formation and evolution in the early stages of the universe. We wait for further improvement in the JWST's measurements of early galactic evolution. Masses of the primordial black holes derived from the scalarons addressing the lithium problem are found to {be within the} mass window that is eligible to act as non-baryonic dark matter. We wish to conclude that a power law gravity theory with primordial black holes of sub-planetary mass has a Big Bang equivalence.

\section*{Acknowledgements}

The corresponding author acknowledges Ranjeev Misra of IUCAA, Pune for his helpful comments during an astronomy meeting that led to the initial development of the idea in this paper. The authors deeply acknowledge the helpful comments of anonymous reviewer.


\bibliography{sample631}{}
\bibliographystyle{aasjournal}

\end{document}